\documentclass{article}
\usepackage[title]{appendix}
\usepackage[utf8]{inputenc}
\usepackage[onehalfspacing]{setspace}
\usepackage{xcolor}
\usepackage{graphicx}

\usepackage{url,hyperref,lineno,microtype,subcaption}
\graphicspath{{figures/}} % Location of the graphics files
\usepackage[natbibapa]{apacite}
\usepackage[section]{placeins} % for floats
\usepackage{listings}
\usepackage{color}
\usepackage{authblk}
\usepackage{float}
\usepackage{ccicons}

\definecolor{light-gray}{gray}{0.95}
\usepackage{lmodern}  % for bold teletype font
\usepackage{amsmath}  % for \hookrightarrow
\usepackage{xcolor}   % for \textcolor
\usepackage{hyperref}
\hypersetup{
    urlcolor=cyan
    }

\usepackage{geometry}
\usepackage{fancyhdr}
\fancypagestyle{plain}{%
\fancyhf{}% clear all header and footer fields
\fancyhead[L]{T.J. M\"oller}
\fancyhead[R]{An Arduino based heartbeat detection device (ArdMob-ECG) for real-time ECG analysis}
\fancyfoot[R]{\textbf{\thepage}} 
}

\onecolumn

\newif\ifdraft
\drafttrue
\ifdraft
    \newcommand{\TODO}[1]{\textcolor{red}{{[\textbf{TODO}: #1]}}}
    \newcommand{\TIM}[1]{\textcolor{violet}{{[\textbf{Tim}: #1]}}}
    \newcommand{\KIM}[1]{\textcolor{blue}{{[\textbf{Kim}: #1]}}}
    
\else
    \newcommand{\TODO}[1]{}
    \newcommand{\TIM}[1]{}
    \newcommand{\KIM}[1]{}
    \renewcommand{\sout}[1]{}
    
    \renewcommand{\textcolor}[2]{#2}
\fi  

\title{An Arduino based heartbeat detection device (ArdMob-ECG) for real-time ECG analysis}
\author[1,2]{Tim Julian M\"oller} 
\author[3]{Martin Voss}
\author[1,2]{Laura Kaltwasser}

\affil[1]{Berlin School of Mind and Brain, Humboldt-Universität zu Berlin, Germany}
\affil[2]{Department for Psychiatry and Psychotherapy, Charité University Medicine, Berlin, Germany}
\affil[3]{Department of Psychiatry and Psychotherapy, Charité University Medicine and St. Hedwig Hospital, Berlin, Germany}   
\begin{document}
\maketitle

% \hspace{-0.5cm}Tim Julian Möller\,$^{1,2*}$\\

% \hspace{-0.5cm}{$^{1}$Berlin School of Mind and Brain, Humboldt-Universität zu Berlin, Germany\\
% $^{2}$ Department for Psychiatry and Psychotherapy, Charité University Medicine, Berlin, Germany}\\
% \vspace{1cm}
% \hspace{-0.65cm} Correspondence$^{*}$\\
% Tim Julian Möller\\
% tim.julian.moeller@gmail.com\\

% \maketitle

% \vspace{1cm}

\vspace{1cm}
\begin{abstract}
\hspace{-0.5cm}
This technical paper provides a tutorial to build a low-cost (10-100 USD) and easy to assemble ECG device (ArdMob-ECG) that can be easily used for a variety of different scientific studies. The advantage of this device is that it automatically stores the data and has a built-in detection algorithm for heartbeats. Compared to a clinical ECG, this device entails a serial interface that can send triggers via USB directly to a computer and software (e.g. Unity, Matlab) with minimal delay due to its architecture. 
Its software and hardware is open-source and publicly available. The performance of the device regarding sensitivity and specificity is comparable to a professional clinical ECG and is assessed in this paper. Due to the open-source software, a variety of different research questions and individual alterations can be adapted using this ECG.
The code as well as the circuit is publicly available and accessible for everyone to promote a better health system in remote areas, Open Science, and to boost scientific progress and the development of new paradigms that ultimately foster innovation. 
\end{abstract}
\thispagestyle{fancy}

\clearpage

\section{Introduction}
\label{intro}
In the recent past, neuroscience research has identified interoception as an important factor contributing to metacognition, empathy, and the experience of the self \citep{fukushima2011association, meessen2016relationship}. Interoceptive signals such as the heart rate, measured with an ECG has thereby been of special interest. These interoceptive signals can be used for a variety of different applications like research studies, modern therapies, but also video games. For example, different projects develop the integration of the heart rate as feedback into virtual environments to develop more immersive experiences for biofeedback in the treatment of ADHD, for the use of more immersive meditation applications, or even to automatically adapt the difficulty and scariness of video games.

However, clinical ECG machines are usually heavy, expensive, and require being operated by trained technical and medical personnel. Utilizing an easy to use, inexpensive, lightweight, and reliable technical device for measuring heart rate and its components would advance the progress in scientific research. In addition, in light of the global Covid-19 pandemic, an easy to use device would also allow the recording of data at the home of the participants with simple instructions and without any additional risk of the physical presence of the scientists during the pandemic.

This paper describes an Arduino-based heartbeat detection device (ArdMob-ECG) that integrates the AD8232 module for real-time ECG analysis. The ArdMob-ECG is easy to assemble, light-weight, small and transportable, inexpensive, yet reliable, and can be used without clinical personnel with simple instructions. This paper is meant as a guide for scientists to independently assemble the device for their own scientific studies. 

% \clearpage

\section{Electrocardiogram (ECG)}
\label{ECG}
\subsection{ECG characteristics}
\pagestyle{plain}
An Electrocardiogramm (ECG) is a widely-used painless and non-invasive technique that is most commonly used to determine heart rate, rhythm, or frequency. The QRS complex, which represents the ventricular depolarization of the heart, is in focus when analyzing heartbeat data. The QRS complex describes the phases around a heartbeat and consists of three waves --- Q, R, and S, and usually lasts between 0.06–0.10 seconds \citep{kashani2005significance}. The Q-wave starts through the depolarization of the interventricular septum. The highest potential is found in the subsequent R-peak which reflects the depolarization of the myocardium of both heart chambers. The R-wave/peak is usually used for calculating the heart rate variability (HRV). The R-peak is followed by a downward deflection which can be referred to as the S-wave. The time between two R-peaks is the interbeat interval (IBI) \citep{christensen2014quantifying}. 

\subsection{Use of medical ECG machines in scientific research}
An ECG can monitor the heart activity of an individual and detect e.g. arrhythmias that can indicate heart problems. This has an immense meaning for monitoring the participants cardiac health. 
Usually a clinical ECG machine which sometimes costs several thousand US-Dollar is needed and due to the cost and the weight of the ECG machine, usually hospitalization is needed for an ECG-assessment. Furthermore, clinical ECG machines usually do not allow changes in their built-in software nor a direct serial interface to other programs.
Therefore, an inexpensive device ($\approx$ 50 \$) that is easy to use, easy to self-assemble, and easy to transport can provide a solution for more remote areas that do not have a health system in range.

\section{An Arduino-based ECG device using the AD8232 module}
\label{the_device}
An Arduino is a programmable single-board microcontroller that comes with its own software integrated development environment (IDE). The software is programmed in an abstraction of C++ and can be loaded via a USB cable to the Arduino board. The board can be equipped with different sensors and modules. 

\begin{figure}[H]
\begin{center}
\includegraphics[width=0.4\textwidth]{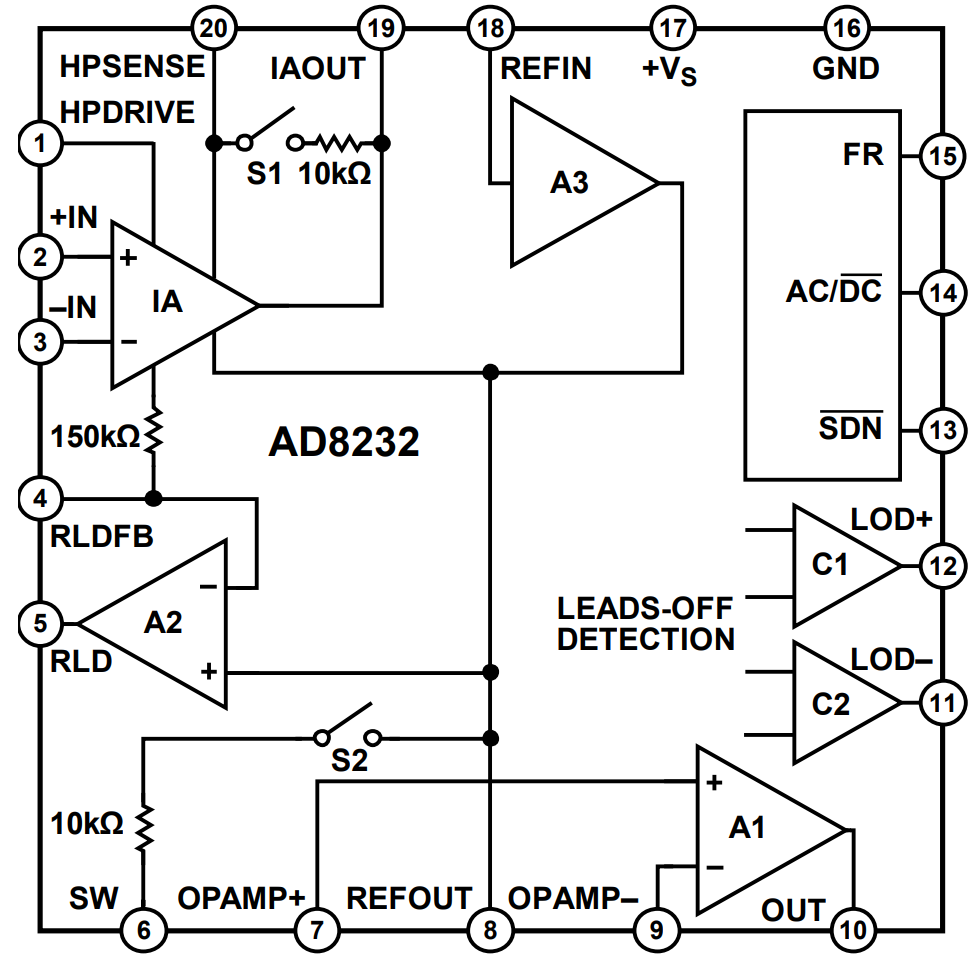}
\caption{A functional block diagram of the AD8232 ECG module taken from \citet{AD8232}}
\label{fig:functional_block_diagram}
\end{center}
\end{figure}

The ArdMob-ECG described in this paper utilizes the AD8232 module, a small chip measuring the electrical signal of the heart that is then translated to an ECG waveform. The AD8232 integrates signals for ECG and other biopotential measurements and is specialized to extract, amplify, and filter small biosignals even in the presence of noisy conditions such as remote electrode placement and movement. The AD8232 consists of specialized instrumentation, operational, and a drive amplifier. Additionally, the chip provides a mid supply reference buffer and a built-in high and low pass filter \citep{kanani2018recognizing}. A schematic functional block diagram of the chip can be seen in \autoref{fig:functional_block_diagram}.

% \vspace{1cm}

The ArdMob-ECG described here is a reliable yet inexpensive, between 10 and 150 Euro, and easy to use ECG device for scientific studies. Due to its light weight and plug and play mechanism, it can also be operated by non-clinical personnel outside of the laboratory. It can also be battery operated without the need of any energy outlets. The device integrates the AD8232 Single Lead Heart Rate monitor module soldered on a standard Arduino board to reliably record data, calculate the heart beat, and save it on an on board micro-SD card.

The ArdMob-ECG implements a simplification of the Pan-Tompkins algorithm to calculate the occurrence of heartbeats in near real time. Different data (e.g. timestamps, raw data) can be saved on a built-in micro SD card, while other data such as triggers can be sent directly to another device using a serial interface (USB-connection). Thus, triggers can be sent with minimal delay. The heartbeat analysis can already be calculated locally on the Arduino.

The Arduino can also provide auditory feedback whenever a heartbeat is detected. This tone can also be easily adjusted to occur faster, slower or with a specific delay to the actual heartbeat, depending on the scientific paradigm (e.g. the widely used interoceptive sensitivity paradigm \citep{garfinkel2015knowing}). 

With respect to scientific purposes, this device can provide solutions for scientific studies since it is fast to compute and introducing a low latency due to a direct objective machine layer (C++) that allows for faster processing, while having a very low power consumption of around 25mA. Furthermore, the variables can be easily modified and additional triggers can be implemented, for example, directly playing the sounds faster or slower according to the last R-R interval \citep{suzuki2013multisensory, brener1993method, wiens2001quadratic}. 

The AD8232 module in combination with an Arduino has been used in prior studies for heartbeat detection \citep[e.g.][]{kanani2018recognizing, bravo2019portable, simanjuntak2020design, bhosale2016healthcare}. However, in order to make it easier for researchers to reconstruct the ArdMob-ECG for use in further studies, the current paper will also provide the code and a clear schematic of the hardware. The following sections provide a comprehensive and detailed description of the device's hardware and software with the aim of enabling researchers to independently assemble the device for use in scientific studies. 
 
\subsection{Design of the Hardware}
\label{hardware}
\begin{figure}[H]
\centering
\includegraphics[width=0.7\textwidth]{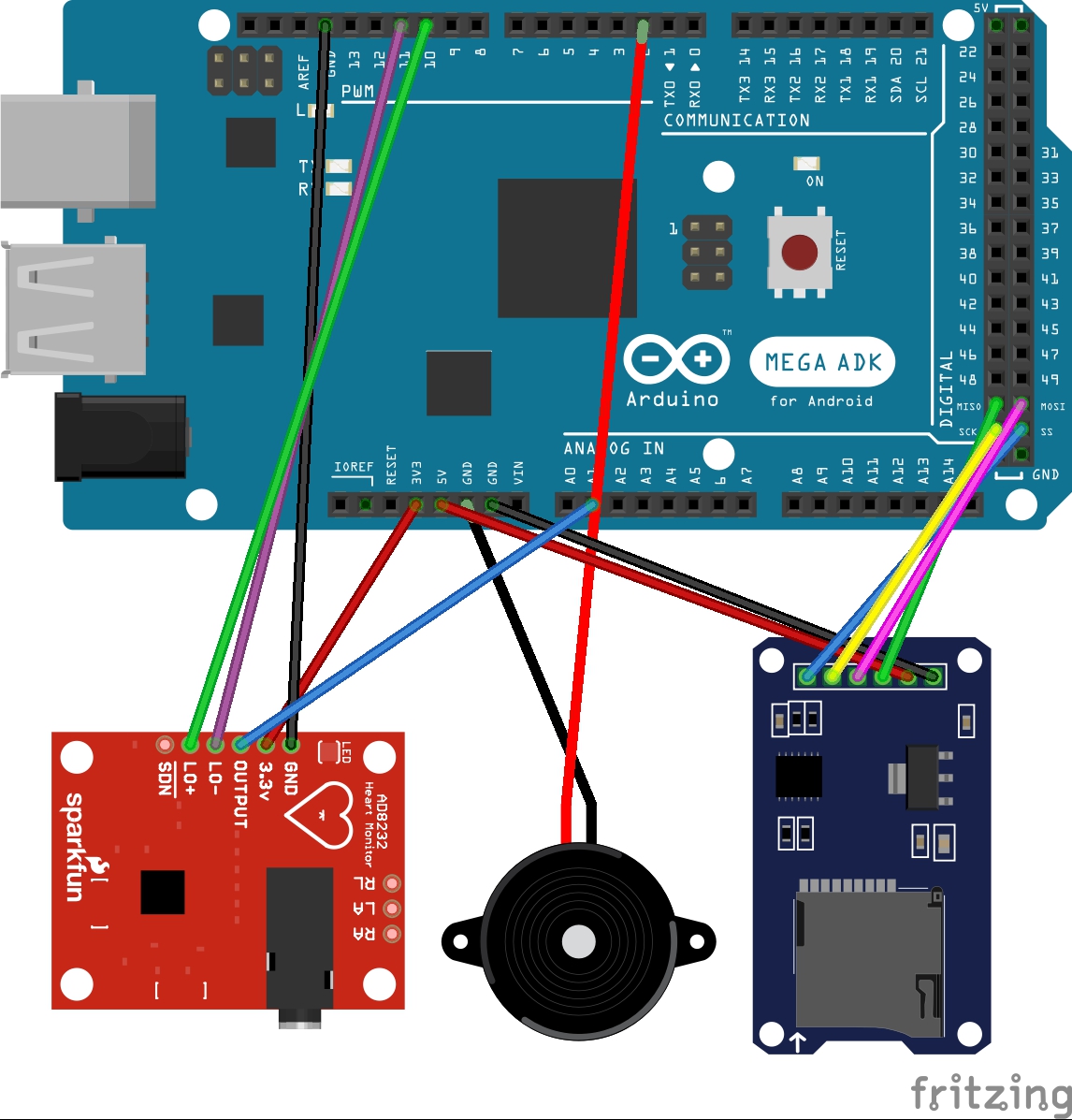}\\
\caption{A schematic diagram of the ArdMob-ECG. The Arduino Mega 2560 board (top, teal) is connected to an AD8232 heart beat monitor (bottom left, red) which has a headphone jack from which the cable for the electrodes connects, to a 85 dB Piezo sound buzzer (bottom center, black) , and a micro SD memory card SPI reader (bottom right, dark blue). (Figure created with Fritzing)}
\label{fig:fritzing_diagram}

\end{figure}

\begin{figure}[H]
\centering
  \begin{minipage}[b]{0.48\textwidth}
\includegraphics[width=1\textwidth]{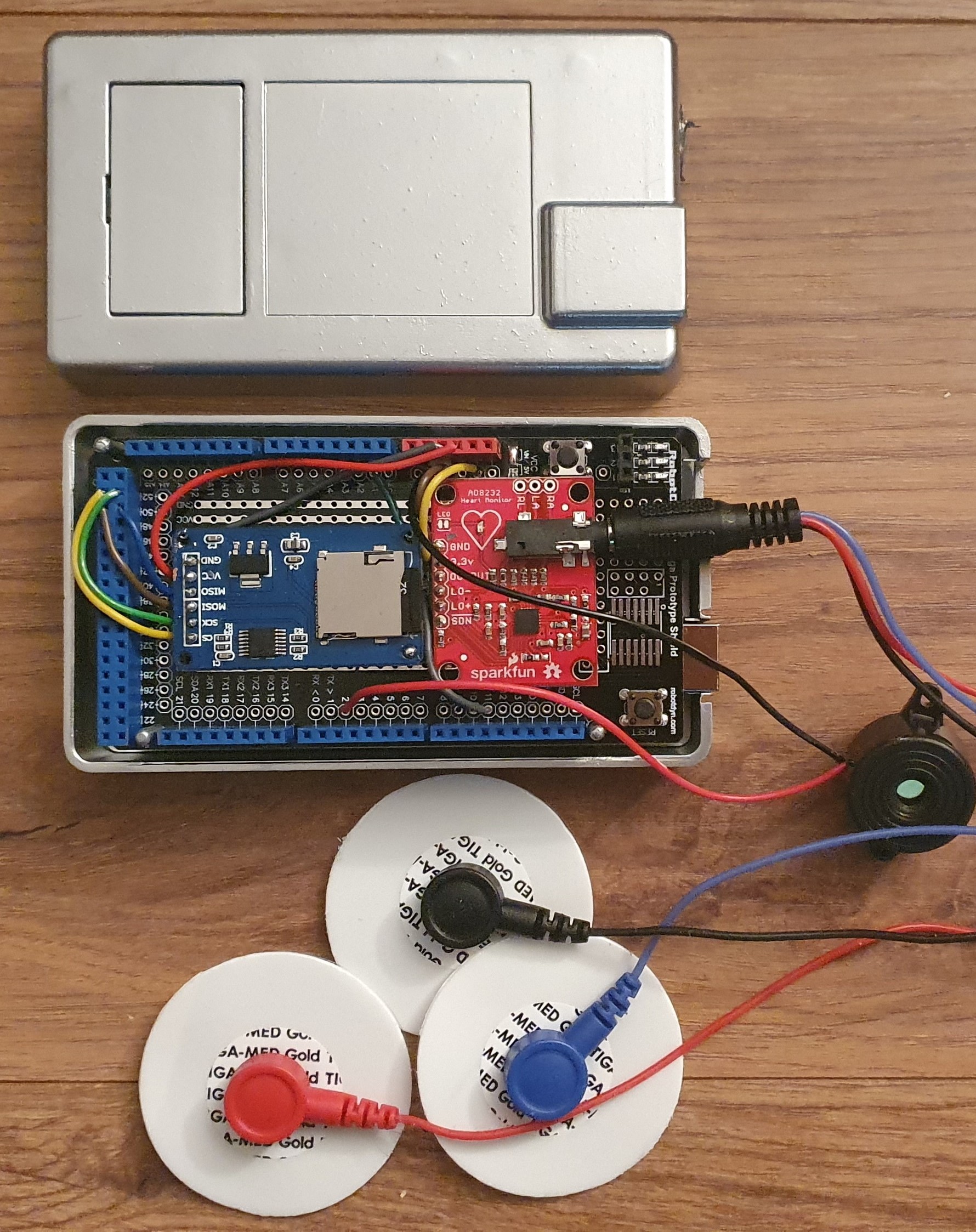}\\
\label{fig:arduino_setup_1}
  \end{minipage}
    \hfill
  \begin{minipage}[b]{0.48\textwidth}
\includegraphics[width=1\textwidth]{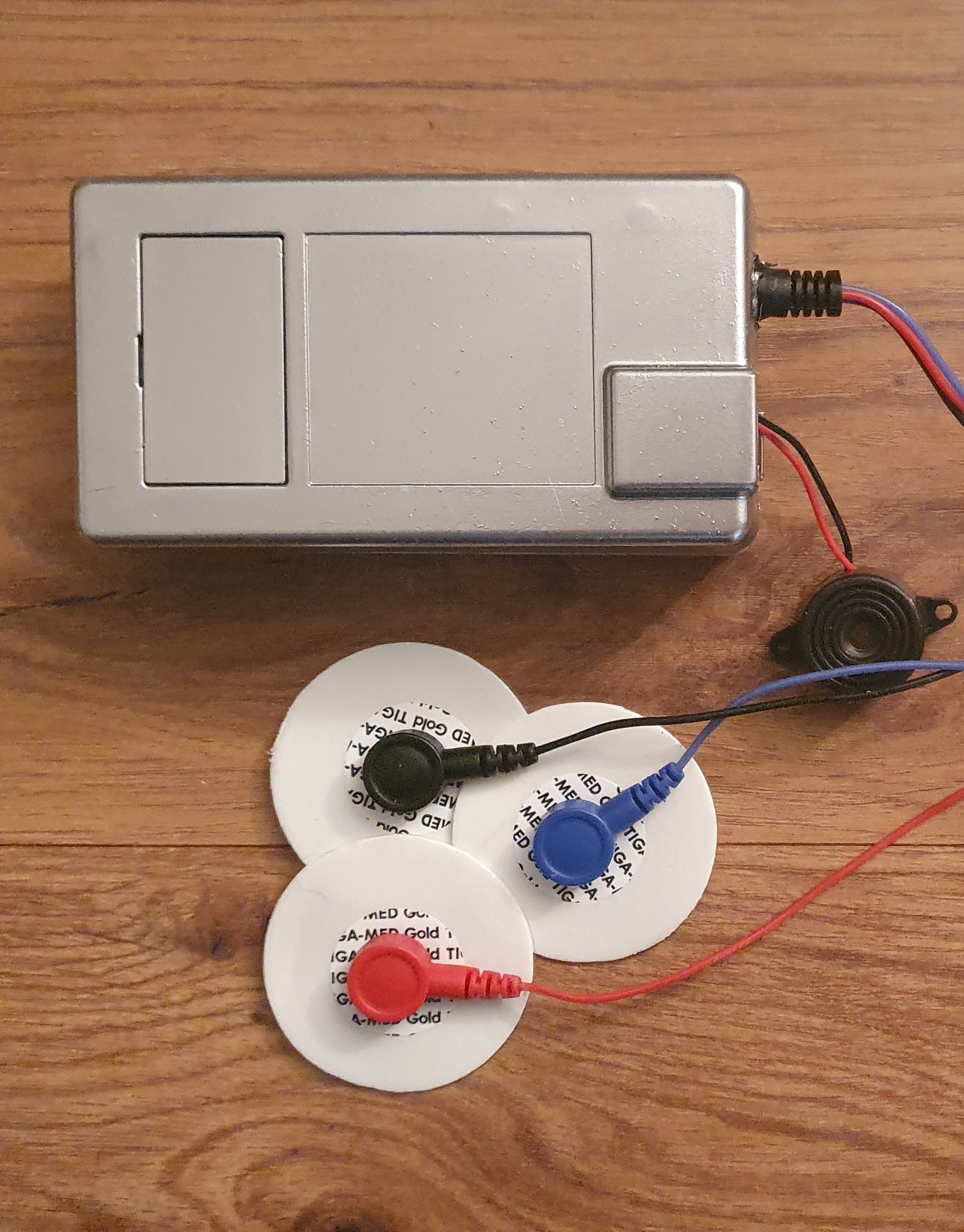}
\label{fig:arduino_setup_2}
  \end{minipage}
  \caption{The ArdMob-ECG inside the aluminium case showing the insides according to the schematic in \autoref{fig:fritzing_diagram} showing the modules and all the relevant cable connections. The cable on the right is the electrodes cable connected to the headphone jack of the AD8232 sensor. On the right is the Piezo sound buzzer. On the left side is the device how it is used.}
\label{fig:arduino_setup}
\end{figure}

The Arduino is capable of a modular build as it can be equipped relatively easy with different sensors, chips, and modules. 
The AD8232 ECG sensor \citep{AD8232} is a module that can be soldered onto an Arduino. It entails a connection cable that is plugged into the 2.54 pin (headphone) jack of the AD8232 module. The connection cable has place for 3 electrodes and can be used with any standard clinical ECG electrodes. 
The electrodes are be attached onto the participant according to the classic lead II configuration which is standard in many scientific studies \citep{christensen2014quantifying, clifford2006advanced}. However, other electrode setups are possible.
An Arduino Mega 2560 R3 with an ATmega2560 processor, equipped with a generic prototype shield was used.
An AD8232 heart-rate module, an electrical piezo sound buzzer (UKCOCO DC 3 – 24 V 85 dB Piezo buzzer), and a micro SD memory card SPI reader (AZDelivery, Deggendorf, Bavaria, Germany) to allow serial communication and on time analysis of the signal were soldered on a standard prototype shield (RobotDyn, Zhuhai, China) that is attached to the Arduino. \autoref{fig:fritzing_diagram} provides a detailed schematic diagram of the circuit.  
Additionally, the ECG-device is embedded in a plastic case that is shielded in aluminium to reduce electrical interference from the surrounding. \autoref{fig:arduino_setup} depicts the setup without the aluminium shielding after soldering the AD232 sensor, and \autoref{fig:arduino_setup} shows the device inside the aluminium case after soldering the Piezo buzzer and with the electrodes cable attached to the headphone jack of the AD8232 sensor. Additional information about practical assembly of the device is presented in \autoref{tips}.

% \vspace{1cm}

\subsection{Design of the Software}
\label{software}
\begin{figure}[H]
\centering
\includegraphics[width=0.7\textwidth]{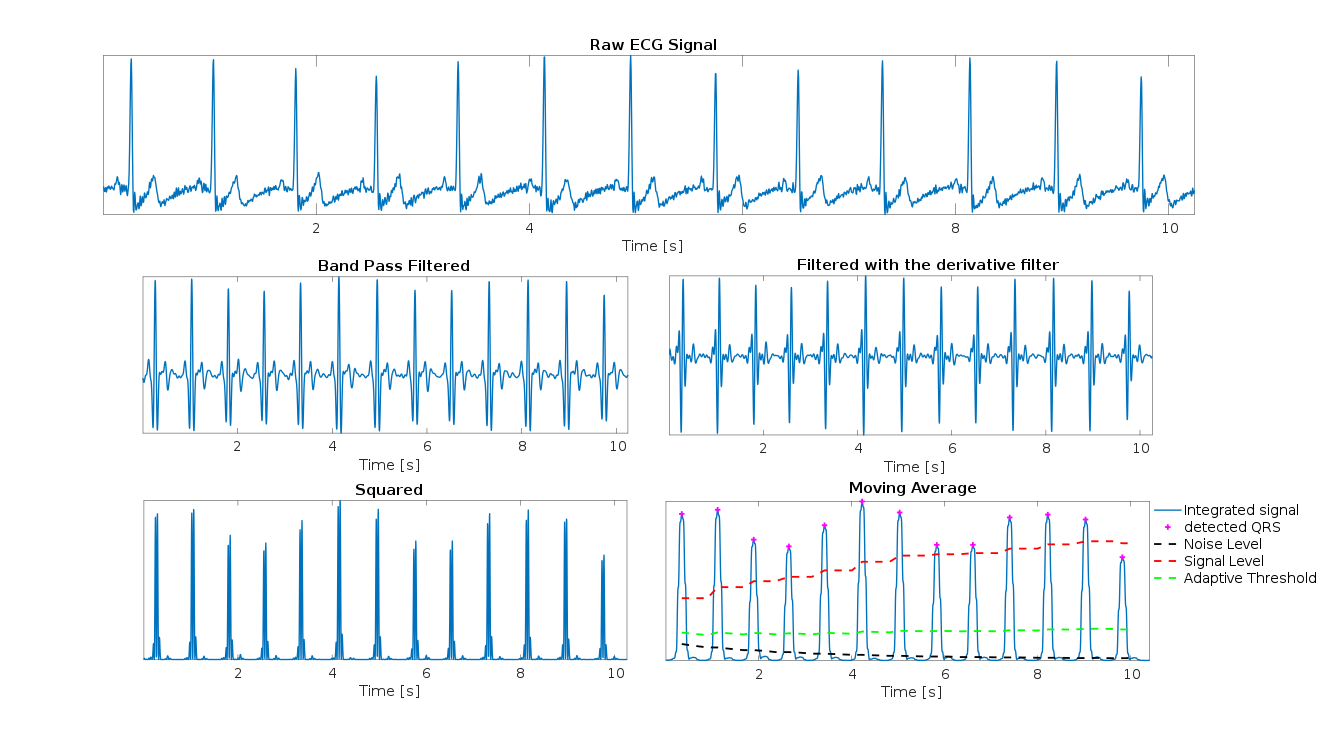}
\caption{The different steps of the Pan-Tompkins algorithm \citep{sedghamiz2014complete, sedghamiz2014matlab, pan1985real}. Taken from \citep{sedghamiz2014matlab}. Note that a high pass filter, followed up by a low pass filter can be seen as a band pass filter, however, doing both steps consecutively saves processing power.}
\label{fig:pan-tompkins}
\end{figure}

In order to analyze an ECG, a peak detection algorithm is usually used. 
One such algorithm is the Pan-Tompkins algorithm, which is commonly used for detecting the QRS complex in ECG, utilizing a series of low and high pass filters to detect the frequencies in the signal and to remove background noise \citep{pan1985real}. Specifically, the Pan-Tompkins algorithm applies an adaptive low pass filter, followed by an adaptive high pass filter in order to reduce background noise. In a next step, this signal is used for an adaptive threshold peak detection (see also \autoref{fig:pan-tompkins}).

Based on the Pan-Tompkins algorithm, the software for the ArdMob-ECG described in this paper was adapted from \citet{milner_real_time_QRS_detection_2015} which poses a simplification of the algorithm used in \citet{chen2003moving}. The software provides an on board real-time analysis of the heartbeat, implementing the adaptive high pass and low pass filtering, and adaptive thresholding of the Pan-Tompkins algorithm. In contrast to the Pan-Tompkins algorithm, no derivative filter and no squaring were implemented. Prerecorded ECG-signals taken from \citet{milner_real_time_QRS_detection_2015} were used to train the adaptive filtering and thresholding models prior to data collection in order to improve the detection accuracy. When noise and muscle artifacts are present that interfere with QRS detection, the algorithm switches back to reliable QRS detection within seconds as soon as the artifacts are absent. In the software, the buffer for the high pass (M) and low pass filter (N) as well as the window size for the QRS-algorithm (winSize) can be tweaked to adjust for the quality of the incoming data. 

The ArdMob-ECG detects and computes the QRS-complex and without the need for further computations, saves it on the microSD card. However, it can also send it directly to a computer or tablet as a trigger. For example, in the VR implementation described in \autoref{implementation}, whenever a heartbeat is detected, a trigger is sent to a PC to activate the Unity code, and the other data is saved on the microSD card. In addition, the code saves the date and timestamp of each data point, and a timer was implemented that counts the elapsed time, as well as a counter for the number of recorded data-points, if enabled. The data is automatically saved on the microSD card and shows an error when no data can be written on the SD card.

With regards to the sound feedback from the Piezo buzzer, it can be decided if tones should be played concurrently with the QRS detection i.e. in the same rhythm, or faster or slower (a percentage can be tuned) with respect to the last QRS-QRS interval.

The software can be found in the code section (\autoref{arduino_code} of this manuscript) as well as in the GitHub repository \citep{moeller_2021}.

% \vspace{1cm}

\subsection{Tips and tricks}
\label{tips}
When first setting up the system, it is recommended to test the equipment for malfunctions first before soldering them together. For the first prototype, it is easiest to connect the modules via jumper cables to the respective pins depicted in \autoref{fig:fritzing_diagram}. The heartbeat detection should already work using the jumper cables even though a lot of noise is introduced. Another testing step before soldering the modules to the Arduino board is to solder them to a prototype shield where all relevant pins are accessible. This will further improve the quality of the data and reduce noise. In this step, make sure that nothing is blocking the slot of the microSD card module. 

Since the micro SD card saves data according to the FAT file system using the 8.3 format, the filename that one chooses for the data must not exceed 8 characters (e.g. the number or the initials of your participant). If there is a file with the same name and format, new data will be appended to the previous datafile. The data can be stored as a .csv or .txt file. To change the format, simply change the extension of the output string accordingly.  

For further information and help with assembly please do not hesitate to contact me via email.

\section{Implementation}
\label{implementation}
The ArdMob-ECG is currently being used in several ongoing studies. In one study, it is used in the interoceptive sensitivity paradigm where participants have to evaluate if they hear their own heartbeat, or if the sound is played faster or slower than their own heartbeat. In a next step, the device detects heartbeats and whenever a heartbeat is detected, a trigger is sent to a PC through a Unity interface to activate a Unity script. In this experiment, depending on the conditions, this leads to the red flashing of a virtual hand in a virtual reality cardiac rubber hand illusion paradigm similar to that described in \citet{suzuki2013multisensory}. The Unity code for this implementation can be found in Appendix B and in the GitHub repository \citep{moeller_2021}. 

\begin{figure}[h]
\centering
  \begin{minipage}[b]{0.48\textwidth}
        \includegraphics[width=01\textwidth]{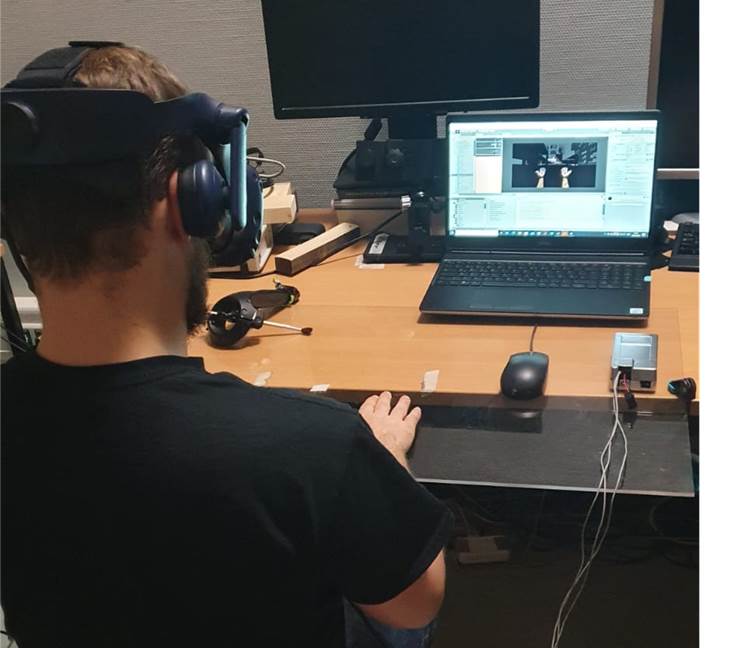}
    \label{Picture_cRHI_Hand_1}
  \end{minipage}
    \hfill
  \begin{minipage}[b]{0.48\textwidth}
    \includegraphics[width=1\textwidth]{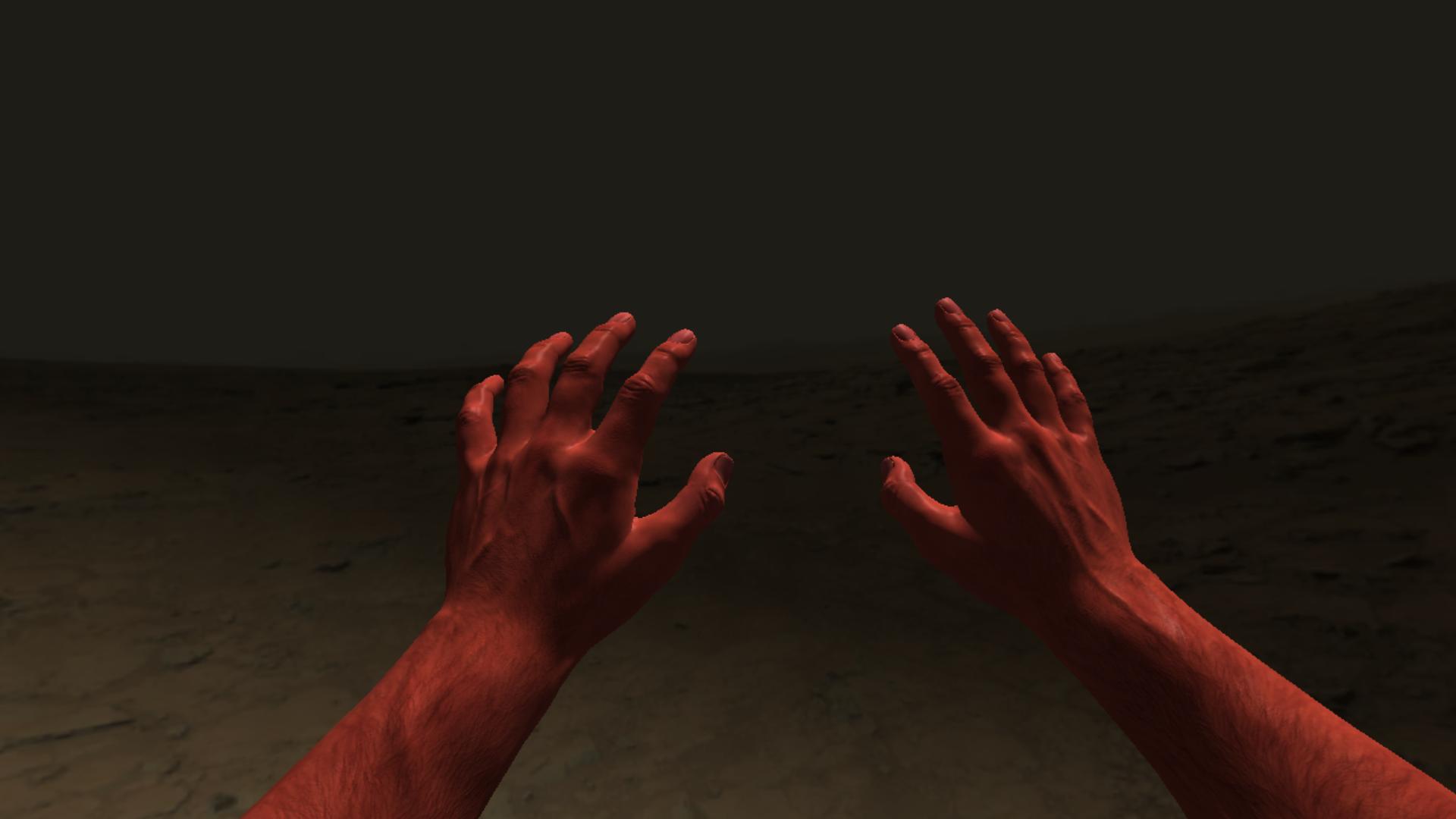}
    \label{Picture_cRHI_Hand_2}
      \end{minipage}
          \caption{A virtual reality study. On the left picture, a participant is doing a virtual reality task, while he is connected to the Arduino-ECG that is to the right. Whenever a heartbeat is detected by the Arduino, his virtual hands flash red (right panel), activated by a calculated R-peak trigger by the Arduino through the serial interface as seen in the left picture.}
\label{Picture_cRHI_Hand}
\end{figure}

\section{Validation}
To assess the quality of data obtained from the three channel ArdMob-ECG, data from 12 channel ECG from a clinical ECG machine (ZOLL X Series CCT Defibrillator) of a German ambulance and was recorded simultaneously for three seconds on one healthy 24 year old participant. The ZOLL X Series CCT Defibrillator ECG-device is standard equipment and is routinely used in german ambulance vehicles. 

\begin{figure}[H]
    \centering
    \includegraphics[width=0.88\textwidth]{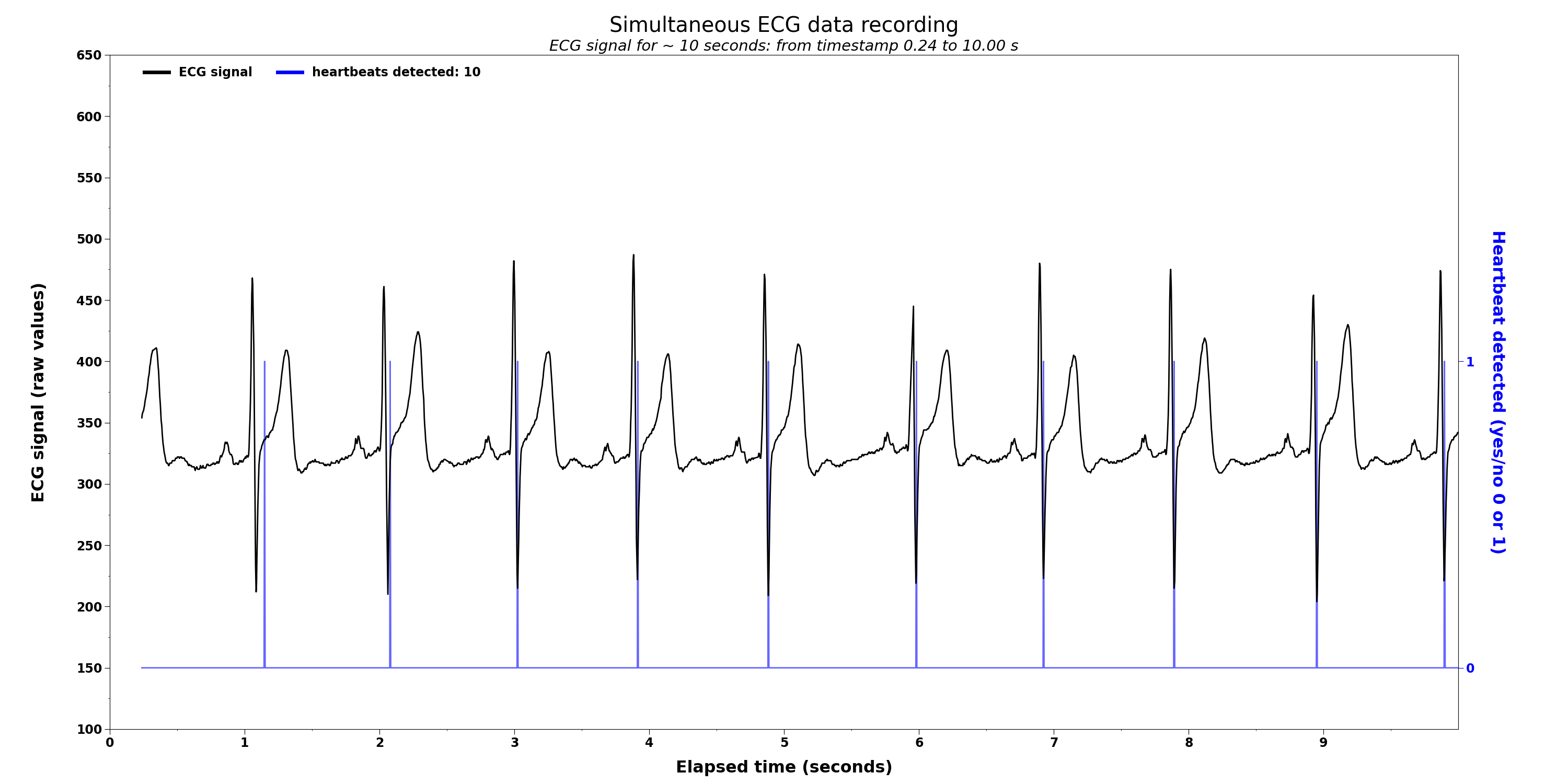}
    \caption{Three electrode ECG data acquired from the ArdMob-ECG, recorded simultaneously with a clinical ZOLL X Series CCT Defibrillator ECG machine (\autoref{Picture_Plot_ECG_professional_recording}). Around 10 seconds of ECG recording. The blue spikes indicate the detection of a heartbeat by the algorithm.}
    \label{Picture_Plot_ECG_data_recording}
\end{figure}

\begin{figure}[H]
\centering
\includegraphics[width=1\textwidth]{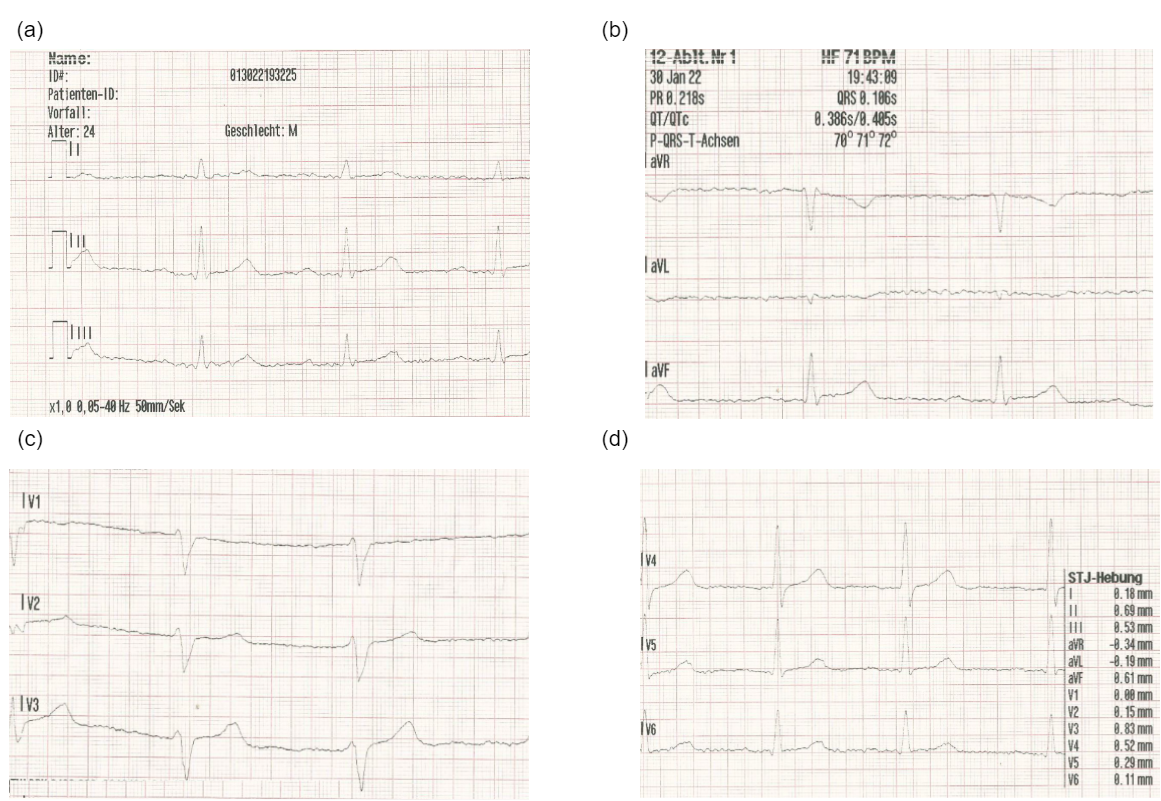}
    \caption{Data from a 12 channel ECG that was received from a professional clinically certified device for comparison. In this image, 3 seconds of an ECG are displayed. Picture a-d all show three seconds of the data of the respective electrode vectors that are depicted to the right in (d).}
    \label{Picture_Plot_ECG_professional_recording}
\end{figure}

\autoref{Picture_Plot_ECG_data_recording} shows the data from the Arduino based ECG device, and \autoref{Picture_Plot_ECG_professional_recording} shows a scan of the data obtained by the clinical ECG machine.  
By visually comparing both Figures, the data quality of the Arduino based ECG seems comparable to the clinical ECG.
Furthermore, a trained medical expert compared both data outputs and assessed the validity of the heartbeat detection. Both devices managed to detect the heartbeats.

\begin{figure}[H]
    \centering
      \begin{minipage}[b]{0.28\textwidth}
    \includegraphics[width=1\textwidth]{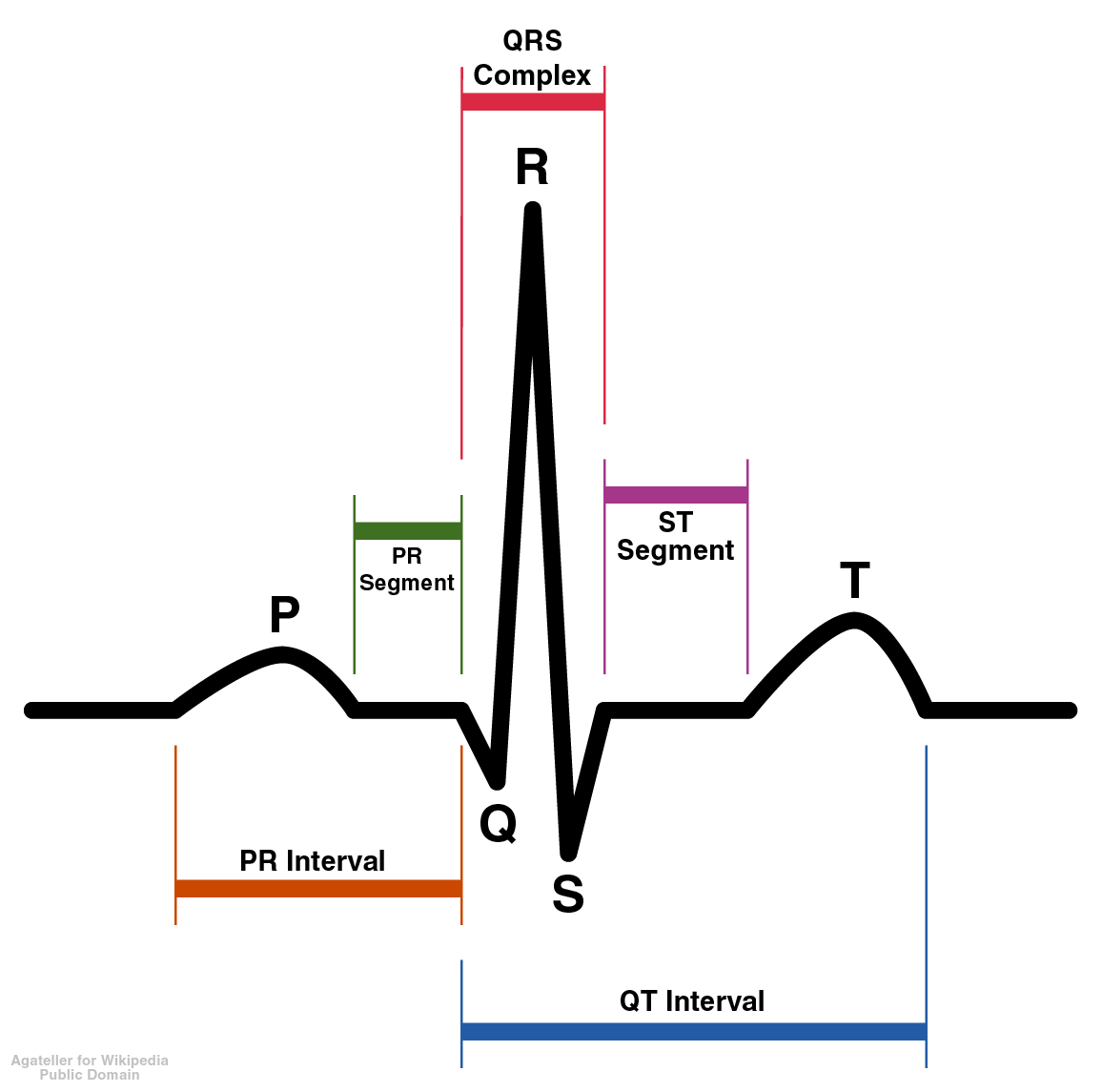}
      \end{minipage}
    \hfill
 \begin{minipage}[b]{0.70\textwidth}
    \includegraphics[width=1\textwidth]{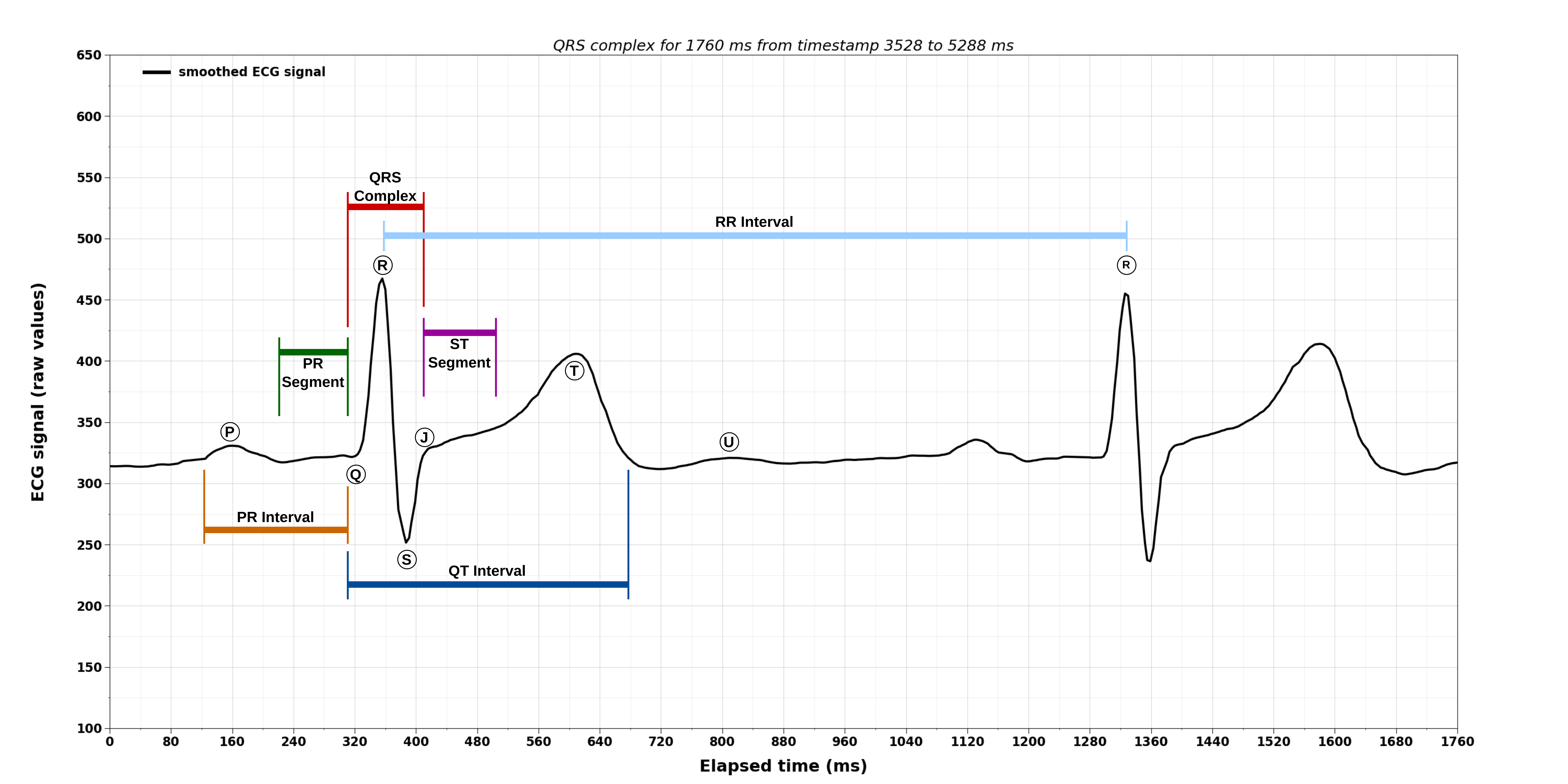}
 \end{minipage}
    \caption{The left panel depicts an ideal schematic representation of a normal ECG consisting of the components P,Q,R,S, and T (Picture taken from \citet{avanzato2020automatic}). The right panel shows a recording from the ArdMob-ECG with the different components.}\label{SinusRhythmLabels.png}
\end{figure}

In addition, a five minutes and three second long ECG using the ArdMob-ECG was recorded. A trained paramedic then examined the recorded ECG signal and counted the heartbeats visually. The software and the paramedic both detected 347 heartbeats within this five minutes and three seconds timeframe. The paramedic also assessed that the ECG detection of the software was always identifying the R-peak correctly. Therefore, the sensitivity and specificity of the Arduino based ECG was 100\%. Besides the R-peak, the paramedic also clearly detected the P,Q,R,S, and T components of the ECG data gathered by the Arduino. 
% We additionally wrote an algorithm that can detect the aforementioned components as well. The code for this detection algorithm can also be found in this paper as well as under the GitHub link. 
\autoref{SinusRhythmLabels.png} shows the P,Q,R,S,T components of a heartbeat in data obtained from the ArdMob-ECG. The raw ECG data was smoothed with a Savitzky–Golay filter.
All in all, the ArdMob-ECG not only reproduces comparable data to the clinical ECG, but also shows sufficient sensitivity and specificity of the heartbeat detection (R-peak) as well as the P,Q,S, and T components. 

\section{Conclusions}
\label{conclusions}
In light of growing development of sophisticated biofeedback methods in the health sciences, ECG data can provide valuable information about internal body states. Currently, the use of an ECG might be limited by expensive, heavy and rather immobile ECG machines, and difficulties in operating them with no technical training which might limit the overall progress in scientific biofeedback studies.

Our mobile ECG device provides a solution to the aforementioned problems of high costs, transportability, and technical training required to operate a clinical ECG machine. Furthermore, the ArdMob-ECG allows for an easy way to save data and to send triggers through a direct interface with minimal temporal delay that can be of use for scientific paradigms.
The open access to the hardware and software allows for an easy reproducibility of the device and its features. It also allows for quick and easy changes for individual requirements, that might not be possible with a clinical ECG. 
In addition, the open access to the software allows for reproducibility of the ArdMob-ECG and quick adaptation for individual studies.

Moreover, another advantage is the possibility to directly communicate with other programs without the need of saving the data or triggers in between on another computer through the serial interface, and the on board processing of the data which prevents an overload of data to other programs. 
This serial interface makes it is easy to use with other experimental or data analysis related software such as Matlab, Python, Blender, or Unity. Different data can be saved on the microSD card. 

The ArdMob-ECG showed good results in several ongoing studies. Also the validation with a medical ECG machine showed no difference in the effective heartbeat detection. 

\clearpage

\section*{Author Contributions}
T.J.M.: Conceptualization, graphical design, writing the original draft \& editing; 
M.V.: Editing; L.K.: Editing
All authors have read and agreed to the published version of the manuscript.

\section*{Conflict of Interest Statement}
The authors declare that the research was conducted in the absence of any commercial or financial relationships that could be construed as a potential conflict of interest.

\section*{Funding}
This research was funded by the Deutsche Forschungsgemeinschaft (DFG, German Research Foundation) – DFG – SPP – 2134

\section*{Usage Notes}
This work is licensed under a Creative Commons Attribution-Noncommercial-ShareAlike License (CC BY-NC-SA). Users are requested to follow the CC BY-NC-SA license, and acknowledge Tim Julian M{\"o}ller in any publication derived from this work, citing this paper as the source. 

The pictures and the code for this device can be found in the GitHub repository \citep{moeller_2021}. The repository is licensed under the MIT license.

\section*{Acknowledgments}
Major sections of the Pan-Tompkins algorithm were adapted from an existing repository \citep{milner_real_time_QRS_detection_2015}.
Some of the code contains elements from other available, open access, existing scripts, described in the Readme file in the GitHub repository. 

Special thanks to Yasmin Kim Georgie for  editing and language improvements that significantly enhanced the quality of this manuscript.

\clearpage%
\bibliographystyle{apacite}
\bibliography{Method_paper}	
\clearpage%

\begin{appendices}
\renewcommand{\thesection}{\Alph{section}.}
\section{Arduino Code}
\label{arduino_code}

\begin{lstlisting}[breaklines=true]
/* Script written by Tim J. Möller
 * Humboldt Universität zu Berlin
 * Berlin School of Mind and Brain, Berlin, Germany
 * Department of Psychiatry and Psychotherapy, Charite University Medicine, Berlin, Germany
 * Contact: tim.julian.moeller@gmail.com
 * 
 * This script records data from the AD8232 shield and plots the data.
 * Preprocessing and detect of the QRS compley is realized through the Pan-Tompkins algorithm
 * (Pan, J., & Tompkins, W. J. (1985). A real-time QRS detection algorithm. IEEE Trans. Biomed. 
 * Eng, 32(3), 230-236.).
 *  This script also saves the Date, the Timestamp, a Timer that counts the passed time and a 
 *  counter for the number of recorded data-points.
 *  
 * Some of the code contains element from available, open access, existing scripts:
 * https://github.com/adafruit/Adafruit_SSD1306
 * https://github.com/blakeMilner/real_time_QRS_detection
 * https://github.com/dxinteractive/ResponsiveAnalogRead
 * SD card read/write (Arduino example sketches) created Nov 2010 by David A. Mellis modified 9 Apr 2012 by Tom Igoe
 * This example code is in the public domain.
 * SD Card test created  28 Mar 2011 by Limor Fried modified 9 Apr 2012 by Tom Igoe
 * This example code is in the public domain.
 * 
 * Last modified: 31.12.2021
 */

  ///////////////// START OF DECLARATION OF IMPORTANT VARIABLES///////////////////////
/*  Here you can find the variables that might need adjustment to you own purpose. A short description what every
 *  variable is doing can be found as a comment after the variable declaration. */
 
const String Filename = "Test_01.txt"; /* Give the Outputfile a name how it will be displayed on the SD Card. Be aware that the
// filename may only consist of a maximum of 8 alphanumerics and has to end with .exe or .csv (e.g. Test_01.txt). If
a file with the same name already exists, this file will be extended. */

bool Plotting = true; /* If you want to plot the data in the Arduino Serial Plotter, set this variable to true. The QRS value
// is then changed to 400 instead of 1, to make the peak more visible. If you use the serial monitor or you want to save the data to
// the SD card, set 'Plotting' to false. */

bool Saving_Command = true; /* If you do not have an SD Card module or USB Card, you cannot save the data. Also, the Sketch
would not compile. In that case, change the Saving_Command to 'false'.*/

bool play_sound = true; /* You can define if you want to activate (true) or deactivate (false) your soundbuzzer. */

bool Experiment_settings = false;

/* If you want to use the sound buzzer, these variables might be relevant for your setup. The Tweakfactor by default is 1.0. 
// If you want to play tones faster or lower compared to the last QRS interval, lower or increase the Tweakfactor. Every 0.1 
// accounts for an increase or decrease of 10% time compared to the original heartbeat.*/
float Tweakfactor = 1.0; // That is the original heartbeat (default)
//float Tweakfactor = 0.7; // Used for playing the tones 30% faster than the actual heartbeat
//float Tweakfactor = 1.3; // Used for playing the tones 30% slower than the actual heartbeat
int tone_freq = 1000; // The frequency of the sound
int tonelength = 20; // The duration of the sound in millisconds

/* Here all the pins are defined*/ 
const int tone_pin =2;
const int heartPin = A1; 
const int SD_chipSelect = 53;

/* Here, the most important parameters for the Pan-Tompkins algorithm can be changed*/
#define M       5 // Here you can change the size for the Highpass filter
#define N       30 // Here you can change the size for the Lowpass filter
#define winSize     250   // this value defines the windowsize which effects the sensitivity of the QRS-detection. 
/* May need adjustments depending on your sample size. If you use a lower sampling rate, a lower windowSize might yield  better results. */
//////////////// END OF DECLARATION OF IMPORTANT VARIABLES///////////////////////

///////////////// Additional libraries and definitions that can be changed if you know what you are doing//////////////////// 
#include <ResponsiveAnalogRead.h> // If you get a compiler error here, you do not have this library. Go to "Tools" --> "Manage libraries" and copy this name in the field on the top. Simply install the library, then you have this library.
#include <SPI.h>// If you get a compiler error here, you do not have this library. Go to "Tools" --> "Manage libraries" and copy this name in the field on the top. Simply install the library, then you have this library.
#include <SD.h>// If you get a compiler error here, you do not have this library. Go to "Tools" --> "Manage libraries" and copy this name in the field on the top. Simply install the library, then you have this library.
#include <Adafruit_GFX.h>// If you get a compiler error here, you do not have this library. Go to "Tools" --> "Manage libraries" and copy this name in the field on the top. Simply install the library, then you have this library.
#define HP_CONSTANT   ((float) 1 / (float) M)
#define MAX_BPM     100 // Set an upper threshold of BPM for a better detection rate
#define RAND_RES 100000000

ResponsiveAnalogRead analog(heartPin, true);
File myFile; 
int saving_interval = 0; //create a variable that counts samples. When an aspired number of samples is gathered, the Arduino will save the data to the .txt file to save computational power
int saving_treshold = 300; //if this treshold is reached, the Arduino saves the gathered data to the output file
float difference = 0;
float interval;

/* Important for calculating the between QRS interval.*/
int cprTimeRead_1 = 0; 
int cprTimeRead_2 = 0;
int timeCPR = 0;
float CPRSUM;

/* Other important variables*/
unsigned long currentMillis = millis();
unsigned long previousMillis = 0;
unsigned long i = 0;
int x=0;
int j=0;
int lastj=0;
int lasty=0;
int LastTime=0;
int ThisTime;
int next_ecg_pt;
int QRS = 0;
int tmp = 0;

/* pre-recorded ECG*/
int s_ecg_idx = 0;
const int S_ECG_SIZE = 226;
const float s_ecg[S_ECG_SIZE] = {1.5984,1.5992,1.5974,1.5996,1.5978,1.5985,1.5992,1.5973,1.5998,1.5976,1.5986,1.5992,1.5972,1.6,1.5973,1.5989,1.5991,1.5969,1.6006,1.5964,1.6,1.5979,1.5994,1.6617,1.7483,1.823,1.8942,1.9581,2.0167,2.0637,2.1034,2.1309,2.1481,2.1679,2.1739,2.1702,2.1543,2.1243,2.0889,2.037,1.982,1.9118,1.8305,1.7532,1.6585,1.6013,1.5979,1.5981,1.5996,1.5972,1.5992,1.599,1.5966,1.6015,1.5952,1.6008,1.5984,1.5953,1.606,1.5841,1.6796,1.9584,2.2559,2.5424,2.835,3.1262,3.4167,3.7061,4.0018,4.2846,4.5852,4.8688,5.1586,5.4686,5.4698,5.1579,4.8687,4.586,4.2833,4.0031,3.7055,3.4164,3.1274,2.8333,2.544,2.2554,1.9572,1.6836,1.5617,1.5143,1.4221,1.3538,1.2791,1.1951,1.1326,1.0407,0.99412,1.0445,1.1262,1.2017,1.2744,1.3545,1.4265,1.5044,1.5787,1.6006,1.5979,1.5988,1.5982,1.5989,1.5982,1.5986,1.5987,1.5983,1.5984,1.5992,1.5965,1.6082,1.6726,1.7553,1.826,1.903,1.9731,2.0407,2.1079,2.166,2.2251,2.2754,2.3215,2.3637,2.396,2.4268,2.4473,2.4627,2.4725,2.4721,2.4692,2.4557,2.4374,2.4131,2.3797,2.3441,2.2988,2.2506,2.1969,2.1365,2.0757,2.0068,1.9384,1.8652,1.7899,1.7157,1.6346,1.5962,1.5997,1.5979,1.5986,1.5989,1.5978,1.5995,1.5976,1.5991,1.5984,1.5981,1.5993,1.5976,1.5993,1.5982,1.5982,1.5993,1.5975,1.5994,1.5981,1.5983,1.5995,1.5967,1.6049,1.6248,1.647,1.664,1.6763,1.6851,1.6851,1.6816,1.6712,1.655,1.6376,1.613,1.599,1.5985,1.5982,1.5989,1.5982,1.5986,1.5987,1.598,1.5991,1.598,1.5987,1.5987,1.598,1.5992,1.5979,1.5988,1.5986,1.598,1.5992,1.5979,1.5988,1.5986,1.598,1.5992,1.5978,1.5989,1.5985,1.5981,1.5992,1.5978,1.599,1.5985,1.5981,1.5992,1.5977,1.599,1.5984,1.5981};

/* timing variables*/
unsigned long previousMicros  = 0;        // will store last time LED was updated
unsigned long foundTimeMicros = 0;        // time at which last QRS was found
unsigned long old_foundTimeMicros = 0;        // time at which QRS before last was found
unsigned long currentMicros   = 0;        // current time
unsigned long RR_peak = 0; 

/* interval at which to take samples and iterate algorithm (microseconds)*/
const long PERIOD = 1000000 / winSize;

#define BPM_BUFFER_SIZE 5
unsigned long bpm_buff[BPM_BUFFER_SIZE] = {0};
int bpm_buff_WR_idx = 0;
int bpm_buff_RD_idx = 0;

/* set up variables to use the SD utility library functions:*/
Sd2Card card;
SdVolume volume;
SdFile root;


void setup() {
    Serial.begin(115200);

    // SD-Card module Setup and error display
      if (Saving_Command == true) {
pinMode(SD_chipSelect, OUTPUT);
    #ifdef SERIAL_USB
    while (!Serial); // wait for Leonardo enumeration, others continue immediately
  #endif

    Serial.print("Initializing SD card...");
  if (!card.init(SPI_HALF_SPEED, SD_chipSelect)) {
    Serial.println("initialization failed. Things to check:");
    Serial.println("* is a card inserted?");
    Serial.println("* is your wiring correct?");
    Serial.println("* did you change the chipSelect pin to match your shield or module?");
    while (1);
  } else {
    Serial.println("Wiring is correct and a card is present.");
  }
  if (!SD.begin(SD_chipSelect)) {
    Serial.println("initialization failed!");
    while (1);
  }
  Serial.println("initialization done.");
myFile = SD.open(Filename, FILE_WRITE);
if (myFile)
      {
        Serial.println("File created successfully.");
        return 1;
      } else
      {
        Serial.println("Error while creating file.");
        return 0;
      }
      
/* Define the labels for your data transfer*/
  myFile.println("Counter,Raw_data,QRS_detected");
  myFile.close();
  myFile = SD.open(Filename, FILE_WRITE);
}
}
void loop() {    
delay(2);
// Update the time, reset the QRS detection and read the next value
  currentMicros = micros();
    previousMicros = currentMicros;
    boolean QRS_detected = false;
      int next_ecg_pt = analogRead(heartPin);
      
////////////////////// WRITE DOWN ALL THE VARIABLES THAT SHOULD BE DISPLAYED HERE ////////////////////////////////////// 
if (Experiment_settings == false){
//Serial.print(i++); // Sample number
//Serial.print(","); 
//Serial.print(millis()); // Sample number
//Serial.print(",");
Serial.print(next_ecg_pt);
Serial.print(","); 
Serial.println(QRS);
//Serial.print(","); 
//Serial.println(RR_peak);
}

/*//////////////////DEFINE WHICH VARIABLES YOU WANT TO TRANSFER TO THE SD CARD EVERY ITERATION. IF YOU WANT TO ADD MORE
// VARIABLES, ADD myFile.PRINT('your datatype'); FOLLOWED BY myFile.print(","); //////////////////////////////////////*/

if (Saving_Command == true) {
if(myFile){
myFile.print(i++); // Sample number
myFile.print(",");
myFile.print(millis()); // Sample number
myFile.print(",");
myFile.print(next_ecg_pt); // Raw ECG data
myFile.print(",");
myFile.print(QRS); //Decision if QRS complex was detected (binary)
myFile.print(","); 
myFile.println(RR_peak);
saving_interval++; // Add a token to the saving interval

/* For performance reasons, the data will be only saved every 300 samples (around every 0.75 seconds). Can be changed by changing the saving threshold
saving_interval++; // Add a token to the saving interval*/
if (saving_interval > saving_treshold){
 saving_interval = 0;
  myFile.flush();
  myFile.close();
myFile = SD.open(Filename, FILE_WRITE);
}
}
}
/* If a QRS is found here, it can be send to an interface e.g. Unity 3D (sendData("QRS")) */

      // give next data point to algorithm
      QRS_detected = detect(next_ecg_pt);
        if(QRS_detected == false) {
        QRS = 0;
      }
      else if(QRS_detected == true){
        foundTimeMicros = micros();
        RR_peak = millis();
        if (Experiment_settings == true){
        Serial.println(QRS);}
        if (Plotting == true) {
        QRS = 400; }
        else if (Plotting == false) {
          QRS = 1; }
        //Serial.println(QRS);
        //delay(10);
        
        sendData("QRS");
        cprTimeRead_2 = cprTimeRead_1;
        cprTimeRead_1 = millis();
        CPRSUM = cprTimeRead_1-cprTimeRead_2;
        interval = CPRSUM*Tweakfactor;
        }

/* Here, the tone will be played according to its Tweakfactor*/
     if((difference > interval) && (play_sound == true)){  
     tone(tone_pin, tone_freq, tonelength); 
     previousMillis = millis();   
  }
  
difference = millis() - previousMillis;
}


/* This section contains the Pan-Tompkins algorithm and is adapted from https://github.com/blakeMilner/real_time_QRS_detection
 Portion pertaining to Pan-Tompkins QRS detection */
// circular buffer for input ecg signal
// we need to keep a history of M + 1 samples for HP filter
float ecg_buff[M + 1] = {0};
int ecg_buff_WR_idx = 0;
int ecg_buff_RD_idx = 0;

// circular buffer for input ecg signal
// we need to keep a history of N+1 samples for LP filter
float hp_buff[N + 1] = {0};
int hp_buff_WR_idx = 0;
int hp_buff_RD_idx = 0;

// LP filter outputs a single point for every input point
// This goes straight to adaptive filtering for eval
float next_eval_pt = 0;

// running sums for HP and LP filters, values shifted in FILO
float hp_sum = 0;
float lp_sum = 0;

// working variables for adaptive thresholding
float treshold = 0;
boolean triggered = false;
int trig_time = 0;
float win_max = 0;
int win_idx = 0;

// number of starting iterations, used determine when moving windows are filled
int number_iter = 0;

boolean detect(float new_ecg_pt) {
        // copy new point into circular buffer, increment index
  ecg_buff[ecg_buff_WR_idx++] = new_ecg_pt;  
  ecg_buff_WR_idx %= (M+1);
 
 
  /* High pass filtering */
  if(number_iter < M){
    // first fill buffer with enough points for HP filter
    hp_sum += ecg_buff[ecg_buff_RD_idx];
    hp_buff[hp_buff_WR_idx] = 0;
  }
  else{
    hp_sum += ecg_buff[ecg_buff_RD_idx];
    
    tmp = ecg_buff_RD_idx - M;
    if(tmp < 0) tmp += M + 1;
    
    hp_sum -= ecg_buff[tmp];
    
    float y1 = 0;
    float y2 = 0;
    
    tmp = (ecg_buff_RD_idx - ((M+1)/2));
    if(tmp < 0) tmp += M + 1;
    
    y2 = ecg_buff[tmp];
    
    y1 = HP_CONSTANT * hp_sum;
    
    hp_buff[hp_buff_WR_idx] = y2 - y1;
  }
  
  // done reading ECG buffer, increment position
  ecg_buff_RD_idx++;
  ecg_buff_RD_idx %= (M+1);
  
  // done writing to HP buffer, increment position
  hp_buff_WR_idx++;
  hp_buff_WR_idx %= (N+1);
  

  /* Low pass filtering */
  
  // shift in new sample from high pass filter
  lp_sum += hp_buff[hp_buff_RD_idx] * hp_buff[hp_buff_RD_idx];
  
  if(number_iter < N){
    // first fill buffer with enough points for LP filter
    next_eval_pt = 0;
    
  }
  else{
    // shift out oldest data point
    tmp = hp_buff_RD_idx - N;
    if(tmp < 0) tmp += (N+1);
    
    lp_sum -= hp_buff[tmp] * hp_buff[tmp];
    
    next_eval_pt = lp_sum;
  }
  
  // done reading HP buffer, increment position
  hp_buff_RD_idx++;
  hp_buff_RD_idx %= (N+1);
  

  /* Adapative thresholding beat detection */
  // set initial threshold        
  if(number_iter < winSize) {
    if(next_eval_pt > treshold) {
      treshold = next_eval_pt;
    }

                number_iter++;
  }
  
  // check if detection hold off period has passed
  if(triggered == true){
    trig_time++;
    
    if(trig_time >= 100){
      triggered = false;
      trig_time = 0;
    }
  }
  
  // find if we have a new max
  if(next_eval_pt > win_max) win_max = next_eval_pt;
  
  // find if we are above adaptive threshold
  if(next_eval_pt > treshold && !triggered) {
    triggered = true;
    return true;
  }
        // else we'll finish the function before returning FALSE,
        // to potentially change threshold
          
  // adjust adaptive threshold using max of signal found 
  // in previous window            
  if(win_idx++ >= winSize){
    // weighting factor for determining the contribution of
    // the current peak value to the threshold adjustment
    float gamma = 0.4;
    
    // forgetting factor - 
    // rate at which we forget old observations
                // choose a random value between 0.01 and 0.1 for this, 
    float alpha = 0.1 + ( ((float) random(0, RAND_RES) / (float) (RAND_RES)) * ((0.1 - 0.01)));
    
                // compute new threshold
    treshold = alpha * gamma * win_max + (1 - alpha) * treshold;
    
    // reset current window index
    win_idx = 0;
    win_max = -10000000;
  }
      
        // return false if we didn't detect a new QRS
  return false;
    
    
}

// When you want to send the data to an interface, it can be done here, but it is possible through the normal Serial.print command, too. 
void sendData(String data){

 #ifdef SERIAL_USB
   Serial.println(QRS); // need a end-line because wrmlh.csharp use readLine method to receive data
   //serialPort.ReadTimeout = 1;
   //delay(1); // Choose your delay having in mind your ReadTimeout in Unity3D
#endif
}
\end{lstlisting}

\clearpage%

\section{Unity Code} 
\label{unity_code}
\begin{lstlisting}[breaklines=true]
using UnityEngine;
using System.Collections;
using System.IO.Ports;

public class Arduino_Read : MonoBehaviour

{
    SerialPort sp = new SerialPort("COM5", 115200);
    public static string QRS;

    void Start()
    {
        sp.Open();
        sp.ReadTimeout = 1;
    }

    void Update()
    {
        try
        {
            QRS = sp.ReadLine();
            print(sp.ReadLine());
        }
        catch (System.Exception)
        {
        }
    }
}
\end{lstlisting}
\end{appendices}
\end{document}